\documentclass[aps,prx,twocolumn,letterpaper,groupedaddress,10pt]{revtex4-2}
\usepackage{amssymb,amsthm,amsmath,amsfonts,mathptmx,mathcmd,mathrsfs}
\usepackage{graphicx,ulem,enumerate,bbm,bm}
\usepackage[pdftex,dvipsnames,usenames]{xcolor}
\usepackage[colorlinks=true,urlcolor=blue,citecolor=blue,linkcolor=blue]{hyperref}
\usepackage{braket,tensor,tikz}

\newcommand{\identity}[0]{\mathbbm 1}

\begin{document}

\title{Resource-efficient universal photonic processor based on time-multiplexed hybrid architectures}

\author{Jonas Lammers$^1$}
    \email{Contact Author: jonas.lammers@uni-paderborn.de}
    \affiliation{$^1$Paderborn University, Integrated Quantum Optics, Institute for Photonic Quantum Systems (PhoQS) Warburger Str. 100, 33098, Paderborn, Germany}
    \affiliation{$^2$Paderborn University, Theoretical Quantum Science, Institute for Photonic Quantum Systems (PhoQS) Warburger Str. 100, 33098, Paderborn, Germany}

\author{Laura Ares$^2$, Federico Pegoraro$^1$, Philip Held$^1$, Benjamin Brecht$^1$, Jan Sperling$^2$}
    \affiliation{$^1$Paderborn University, Integrated Quantum Optics, Institute for Photonic Quantum Systems (PhoQS) Warburger Str. 100, 33098, Paderborn, Germany}
    \affiliation{$^2$Paderborn University, Theoretical Quantum Science, Institute for Photonic Quantum Systems (PhoQS) Warburger Str. 100, 33098, Paderborn, Germany}

\author{Christine Silberhorn$^1$}
    \affiliation{$^1$Paderborn University, Integrated Quantum Optics, Institute for Photonic Quantum Systems (PhoQS) Warburger Str. 100, 33098, Paderborn, Germany}
    \affiliation{$^2$Paderborn University, Theoretical Quantum Science, Institute for Photonic Quantum Systems (PhoQS) Warburger Str. 100, 33098, Paderborn, Germany}

\date{\today}

\begin{abstract}

For the ever-growing field of quantum information processing, large-scale, efficient multi-port interferometers serving as photonic processors are required.
In this context, the suitability of quantum walks as the interferometric base for universal computation has been theoretically proven.
In this work, we bridge the gap between theoretical proposals and state-of-the-art experimental capabilities by providing the recipe for the implementation of a universal photonic processor in discrete-time quantum walks.
Specifically, we present the protocol how to translate arbitrary linear transformations into the coin and step operator of a quantum walk and map these to the experimental parameters of the established time-multiplexed platform \cite{PhysRevLett.104.050502}.
We show that our interface is highly scalable and resource-efficient due to the hybrid encoding consisting of multiple degrees of freedom.
Finally, we prove that our system is highly resilient against experimental imperfections and show that it compares favorably against existing architectures.

\end{abstract}

\maketitle

\section{Introduction}
\label{sec:Introduction}
In the field of optical quantum information processing, most protocols rely on the interference between photons in a multiport interferometer. 
Universal interferometers, capable of implementing any linear transformation between a large number of optical modes, also referred to as photonic  processors, are at the heart of photonic quantum computation schemes \cite{Briegel2009, PRXQuantum.2.030325, doi:10.1126/science.aab3642}.
They are also powerful tools for optical networking and communication \cite{6008516, Jacobs_2016, PhysRevLett.117.210501}, forming the bases for distributed quantum computation based on hybrid systems \cite{Xu2022}.
With the recent development around boson sampling and Gaussian boson sampling \cite{Zhong2020, doi:10.1126/science.abe8770, Madsen2022, Yu2023, Gao2022}, the scaling of photonic processors towards a larger number of modes has become an important issue. 
To be useful for most applications, photonic processors need to fulfill three criteria: exhibit low loss, involve a high number of optical modes, and implement arbitrary unitary transformations with high fidelity.

In recent years, multiple platforms for realizing photonic processors have emerged.
The most prevalent platform utilizes spatial encoding, where photonic processors are based on a network of beam splitters and phase shifters following either the configuration by Reck et. al. \cite{PhysRevLett.73.58} or Clements et. al. \cite{Clements:16}. 
While these schemes have shown high fidelity and low loss in integrated photonic chips \cite{Hoch2024, barzaghi2025lowloss24modelaserwrittenuniversal, Arrazola2021}, the number of components required scales quadratically with the number of modes.
In order to overcome this scaling problem, research has shifted towards using different encoding schemes, resulting in two separate approaches.
The first approach is based on time-frequency-bin encoding \cite{PhysRevResearch.6.L022040, Serino:25, PRXQuantum.5.040329}, allowing for monolithic implementations that reduce the number of required components to one. 
Monolithic photonic frequency processors are generally based on non-linear conversion processes, which suffer in current state of the art demonstration from low efficiencies and requiring high spectral resolution \cite{PRXQuantum.4.020306}.
The second approach to reduce the number of components is to reuse them in a loop architecture, where the underlying mode structure is encoded in time bins \cite{PhysRevLett.104.050502, PhysRevLett.113.120501, PhysRevA.91.012306}. 
Using protocols based purely on time bins allows for non-universal photonic processors with extremely high number of modes and low losses \cite{Madsen2022}, as well as for universal systems with high losses \cite{Yu2023}.
A promising way to solve this tradeoff between universality and loss is time-multiplexed hybrid systems combining multiple degrees of freedom in a time-multiplexed architecture.
Its feasibility has been theoretically and experimentally pointed out, for instance, in Refs. \cite{PhysRevLett.104.050502, PhysRevA.99.062301, borghi2025quantuminterferencetimefrequencymodes}.

In the adjacent field of quantum simulation, generalized hybrid systems, namely quantum walks, have been a long-established platform based on large-scale interferometry. 
Quantum walks, as originally developed by Y. Aharonov \cite{QW}, analyze the spread of one or more photons ---typically referred to as walkers --- over a large graph structure. 
They are hybrid systems where one subsystem --- typically referred to as position --- represents the graph's vertices while the other --- typically referred to as coin --- controls the transitions between vertices.
Using the underlying graph representation, it has been proven that quantum walks can serve as a platform for gate-based universal computation \cite{Childs09, Lovett10}.
Furthermore, combined with measurement-induced nonlinearities and single photons, quantum walks are capable of implementing both single-qubit and two-qubit gates in the form of C-Not gates \cite{pegoraro2024}.

As the aforementioned protocols for universal quantum computation strive to implement quantum gates between qubits encoded using the graph's vertices, they neglect the coin subsystem for mode encoding, which worsens scalability.
Furthermore, for a universal photonic processor, we require a protocol (typically referred to as the compiler protocol) capable of translating arbitrary linear transformations $\hat{U}_{\mathrm{Target}}$ into the individual photonic processor parameters that should be programmed.
To our knowledge, there is currently no compiler protocol 
that allows the use of discrete-time quantum walks as a universal processor in such a way that both the coin and position degrees of freedom are used to encode the underlying mode structure.

In this work, we show how a discrete-time quantum walk can be directly used as a universal photonic processor.
We provide the compiler protocol, translating arbitrary linear transformations into their smallest footprint in a quantum walk.
Furthermore, we introduce a setup scheme for quantum walks in the time-multiplexed domain that can perform as universal processors by applying our findings.
We prove that, in our scheme, transition probabilities are completely immune to most expected experimental causes of losses and phase noise.
Finally, we carry out a systematic study on the influence of experimental imperfections, comparing the performance of the different state-of-the-art platforms for universal processors.
We conclude that the proposed setup compares favorably with existing schemes in terms of loss and phase noise resilience.

\section{Quantum Walk}
\label{sec:Architecture}

Discrete-time quantum walks (DTQWs) describe the evolution of one or more quantum particles on a graph structure.
Each walker starts localized at some graph vertex 
and its probability amplitude moves to one of the adjacent vertices according to its coin state, allowing for quantum interference.
The state of the walker is therefore composed of its position and coin state, belonging in the composite Hilbert space $\mathcal{H} = \mathcal{H}_\mathrm{pos} \otimes \mathcal{H}_\mathrm{coin}$.
Here, we consider a two-dimensional coin-space, $\mathcal{H}_\mathrm{coin} = \mathrm{span}\left\{\ket{0},\ket{1} \right\}$, and an infinite-dimensional position Hilbert space $\mathcal{H}_\mathrm{pos} = \mathrm{span} \left\{\ket{x} : x \in \mathbb{Z} \right\}$.

The evolution of a DTQW occurs by iteratively manipulating the coin state of the walker, followed by the transition towards the next position according to its coin state.
Mathematically, we describe each iteration $n$ as performing a coin operation followed by a shift operation.
The coin operation manipulates the coin state of the walker at each position and each iteration individually,
\begin{equation}
\label{CoinOperator}
    \hat{C}(n) = \sum_{x \in \mathbb{Z}} \ket{x}\bra{x} \otimes
        \sum_{p,q=0}^{1} c_{p,q}^{(x)}(n) \ket{p}\bra{q},
\end{equation}
where $c_{p,q}^{(x)}$ represents the transition probability amplitude from the coin state $\ket{q}$ to $\ket{p}$, at position $x$.
The shift operator transforms the position of the walker as
\begin{equation}
\label{ShiftOperator}
    \hat{S} = \sum_{x \in \mathbb{Z}} \left(
            \ket{x-1}\bra{x} \otimes \ket{0}\bra{0}+ \ket{x + 1}\bra{x} \otimes \ket{1}\bra{1}
        \right).
\end{equation}
Here we construct the unitary at each step $\hat{U}(n)$ as firstly applying the shift and then the coin operation, as this matches the proposed experimental setup more closely. Note that the following treatment can be done equivalently for a unitary operator that firstly applies the coin and then the shift operation.
Thus, the unitary evolution at each step is $\hat{U}(n) = \hat{C}(n)\hat{S}$, which, after $N$ steps, generates the full evolution of the quantum walk as
\begin{equation}
    \label{eq:u(n)}
    \hat{U}_N=\prod_{n=1}^N \hat{U}(n).
\end{equation}

The aim of this work is to implement any target unitary by a discrete-time quantum walk $\hat{U}_\textrm{Target}=\hat{U}_N$, where every step $\hat{U}(n)$ in Eq. \eqref{eq:u(n)} corresponds to one loop of our proposed experimental setup presented in Section \ref{sec:Setup}.

\section{Building-block decomposition}
\label{sec:Decomposition}

In this section, we present the mathematical decomposition of the unitary evolution $\hat{U}_N$ that allows us to find the configuration for the experimental setup (cf. Sec. \ref{sec:Setup}) that will implement the target unitary $U_\mathrm{Target}$.
First, we identify the smallest unit cell of a coined quantum walk evolution such that the complete process can be represented as an iteration of this element.
Next, we map this unit cell to a local network representation where it can be understood as applying a specific sequence of next-neighbor operations.
Using the derived local network representation, we utilize the decomposition proposed in Ref. \cite{debrugière2025newdesignslinearoptical} to show how to determine the individual transition amplitudes of the coin operator to realize arbitrary target unitaries.
In total, this section lays the theoretical fundamentals for the compiler algorithm presented in Section \ref{sec:CompilerProtocol}.

\subsection{Unit cell}

We first decompose the quantum walk position space into even and odd components, $\mathcal{H}_\mathrm{pos, e} = \mathrm{span}\{\ket{x} : x \in 2 \mathbb{Z}\}$ and $\mathcal{H}_\mathrm{pos, o} = \mathrm{span}\{\ket{x} : x \in 2 \mathbb{Z} + 1\}$, respectively.
Thus, the complete Hilbert space can be written as $\mathcal{H} = \left(\mathcal{H}_\mathrm{pos, e} \oplus \mathcal{H}_\mathrm{pos, o}\right) \otimes \mathcal{H}_\mathrm{coin} = \mathcal{H}_\mathrm{e} \oplus \mathcal{H}_\mathrm{o}$, where $\mathcal{H}_\mathrm{e/o}=\mathcal{H}_\mathrm{pos, e/o} \otimes \mathcal{H}_\mathrm{coin}$ and $\oplus$ denotes the direct sum.

Following this decomposition, we can group even and odd positions, $\{-N,\dots,\-2,0,2,\dots,N,-N+1,\dots,-1,1,\dots,N-1\}$, and rewrite the coin operator in a matrix form,
\begin{equation}
    \hat{C}(n) = 
        \begin{pmatrix}
            \hat{C}_e(n) & 0 \\
            0 & \hat{C}_o(n)
        \end{pmatrix},
\end{equation}
where $\hat{C}_\mathrm{e/o}(n)$ acts on $\mathcal{H}_\mathrm{e/o}$.
The shift operator switches even to odd positions and vice versa, which generates the anti-diagonal block structure
\begin{equation}
    \hat{S} =
        \begin{pmatrix}
            0 & \hat{S}_{o} \\
            \hat{S}_{e} & 0
        \end{pmatrix}.
\end{equation}
In this matrix form, the evolution operator for each step reads as 
\begin{equation}
    \hat{U}(n) = \begin{pmatrix}0 & \hat{C}_e(n)\hat{S}_{o} \\ \hat{C}_o(n)\hat{S}_{e} & 0\end{pmatrix},
\end{equation}
which clearly shows that even and odd positions evolve independently. 
That is, each step switches the space of positions we are working with.
To avoid this, we directly consider a two-step evolution,
\begin{equation}
\label{Eq:U2}
\begin{split}
    &\hat{U}(n + 1) \hat{U}(n)\\ 
    =& \begin{pmatrix} 
        \hat{C}_e(n + 1)\hat{S}_{o}\hat{C}_o(n) \hat{S}_{e} & 0 \\
        0 & \hat{C}_o(n + 1)\hat{S}_{e} \hat{C}_e(n)\hat{S}_{o}
    \end{pmatrix},
\end{split} 
\end{equation}
where even positions stay even and odd positions stay odd, completely dividing the operation space.
It is therefore sufficient to study the structure of the linear transformation of a double step in one position subspace.
Note that, for an odd number of steps, we can still define the unit-cell until the final step, followed by a switch between even and odd positions that can be effectively implemented by a permutation operation.
The resulting unit-cell for the evolution of even positions in any $N$-step quantum walk is therefore given by $\hat{U}_{\mathrm{cell}}(n) = \hat{C}_e(n + 1) \hat{S}_{o} \hat{C}_o(n) \hat{S}_{e}$. 

Furthermore, note that as even and odd positions evolve independently in the quantum walk network, they can be used to implement two parallel unitary transformations.
This could be useful as this allows us to, for example, implement indefinite casual order experiments \cite{PhysRevLett.113.250402}, through a straightforward population of both even and odd positions and interfering them after the quantum walk network.

While the unit-cell operation only transforms even-numbered positions into even-numbered positions, as seen in Eq. \eqref{Eq:U2}, $\hat{S}_{e}$ and $\hat{S}_{o}$ still switch even and odd positions.
The next step is to transform these operators such that each operator of the individual blocks exclusively acts on one subspace.
To this end, we introduce the so-called transition operator 
\begin{equation}
    \hat{M} = \sum_{x \in \mathbb{Z}} \ket{x + 1}\bra{x} \otimes \hat{\mathbbm 1}_{\mathrm{coin}},
\end{equation}
where $\hat{\mathbbm 1}_{\mathrm{coin}}$ is the identity matrix acting on the coin subsystem. 
This hermitian operator allows us to define operators which all act on the same subspace, $\mathcal{H}_e$,
\begin{equation}
    \hat{U}_{\mathrm{cell}}(n) =
    \underbrace{\hat{C}_e(n + 1)}_{\hat{C}_e(n + 1)}
    \underbrace{\hat{S}_{o}\hat{M}}_{\hat{S}_{o}^{'}} 
    \underbrace{\hat{M}^\dagger \hat{C}_o(n) \hat{M}}_{\hat{C}^{'}_o(n)}
    \underbrace{\hat{M}^\dagger \hat{S}_{e}}_{\hat{S}_{e}^{'}}.
\end{equation}
In summary, we have constructed a two-step unit-cell of an $N$-step evolution, which acts and is comprised of operators that only act on the even-numbered position space $\mathcal{H}_e$. The next step is to find the underlying structure of this unit cell that enables us to map an $N$-step evolution to a local network.

\subsection{Local network representation}
As we have seen above, the space of states in a coined quantum walk has a bipartite structure where we can identify coin and position degrees of freedom as the subsystems.
Once we restrict to the subspace of even positions, we introduce the bijective function $f: 2\mathbb{Z} \otimes \{0, 1\} \rightarrow \mathbb{Z}, (x, p) \stackrel{f}{\mapsto} x + p$, serving as an isomorphism between spaces.

Working in the transformed Hilbert space $f(\mathcal{H}_e)$, we now introduce a new transition operation ${\hat{M}^{'}} = \sum_{z \in \mathbb{Z}} \ket{z}\bra{z + 1}$ and insert it into our unit cell as follows:
\begin{equation}
    \hat{U}_{\mathrm{cell}}(n) =
    \hat{C}_e(n + 1)
    \underbrace{\hat{S}_{o}^{'}\hat{M}^{'}}_{\hat{S}_{o}^{''}} 
    \underbrace{(\hat{M}^{'})^\dagger \hat{C}_{o}^{'}(n) \hat{M}^{'}}_{\hat{C}_{o}^{''}(n)}
    \underbrace{(\hat{M}^{'})^\dagger \hat{S}_{e}^{'}}_{\hat{S}_{e}^{''}}.
\end{equation}
The resulting operators, representing our unit cell, can now be written as 
\begin{equation}
    \begin{aligned}
        \hat{C}_e(n)
            &= \sum_{x \in 2\mathbb{Z}} \sum_{p, q = 0}^1 c_{p, q}^{(x)}(n) \ket{x + p}\bra{x + q},\\
        \hat{C}_o^{''}(n) 
            &= \sum_{x \in 2\mathbb{Z}} \sum_{p, q = 0}^1 c_{p, q}^{(x + 1)}(n) \ket{x + 1 + p}\bra{x + 1 + q},\\
        \hat{S}_{o}^{''}
            &= \sum_{x \in 2\mathbb{Z}} \sum_{p = 0}^1 \ket{x + 2 - p}\bra{x + 1 + p},\\
        \hat{S}_{e}^{''}
            &= \sum_{x \in 2\mathbb{Z}} \sum_{p = 0}^1 \ket{x + 1 - p}\bra{x + p},\\
    \end{aligned}
\end{equation}
where all operations act only on neighboring modes, resulting in a local network. 
In order to better understand the structure of this local network, we decompose the coin operators into a sequence of beam-splitter operators acting on modes $a$ and $b$,
\begin{equation}
    \begin{aligned}
        \hat{T}_{a,b}(n) 
            &= \sum_{z \in \mathbb{Z}\backslash\{a, b\}} \ket{z}\bra{z} + \sum_{i, j \in \{a, b\}} T_{a,b}^{(i,j)}(n) \ket{i}\bra{j}\\
            &= \begin{pmatrix}
                \ddots & \vdots & \vdots & \vdots & \vdots & \\
                \cdots & 1 & 0 & 0 & 0 & \cdots\\
                \cdots & 0 & T_{a,b}^{(a,a)}(n) & T_{a,b}^{(a,b)}(n) & 0 & \cdots\\
                \cdots & 0 & T_{a,b}^{(b,a)}(n) & T_{a,b}^{(b,b)}(n) & 0 & \cdots\\
                \cdots & 0 & 0 & 0 & 1 & \cdots\\
                 & \vdots & \vdots & \vdots & \vdots & \ddots\\
            \end{pmatrix}
    \end{aligned},
\end{equation}
with $b = a + 1$ and $a \in \mathbb{Z}$. 
This allows us to rewrite the coin operators as

\begin{equation}
    \begin{aligned}
        &\hat{C}^{''}_e(n) = \prod_{x \in 2\mathbb{Z}} \hat{T}_{x, x + 1}(n) \quad \mathrm{and}\\
        &\hat{C}_o^{''}(n) = \prod_{x \in 2\mathbb{Z}} \hat{T}_{x + 1, x + 2}(n),
    \end{aligned}
\end{equation}
with $T_{a,a + 1}^{(i,j)}(n) = c_{i - a, j - a}^{(a)}(n)$.

Besides, the shift operators are permutation operations, where, $\forall x \in 2\mathbb{Z}$, $\hat{S}_{o}^{''}$ routes $x + 1 \leftrightarrow x + 2$ , while $\hat{S}_{e}^{''}$ routes $ x \leftrightarrow x + 1 $.
We can also represent these shift operations as next-neighbor beam splitters 
\begin{equation}
    \begin{aligned}
        &\hat{S}^{''}_e = \prod_{x \in 2\mathbb{Z}} \hat{T}_{x, x + 1}(n)\quad\mathrm{and} \\
        &\hat{S}^{''}_o = \prod_{x \in 2\mathbb{Z}} \hat{T}_{x + 1, x + 2}(n),
    \end{aligned}
\end{equation}
where each beam splitter implements a Pauli-X gate.
This further implies that they can be merged with the beam-splitter operations of the coin operators, changing the individual weights without changing the structure of the interferometer.

In total, we decompose the unit cell operation as
\begin{equation}
    \hat{U}_{\mathrm{cell}}(n) = \prod_{x \in 2\mathbb{Z}} \hat{T}_{x, x + 1}(n + 1) \prod_{y \in 2\mathbb{Z}} \hat{T}_{y + 1, y + 2}(n),
\end{equation}
implementing a local network with an infinite dimensionality.

Finally, we introduce reflective boundary conditions to map the infinite-dimensional quantum walk evolution $\hat{U}_N$, acting on $f(\mathcal{H}_{e})$, to the target unitary $\hat{U}_{\mathrm{Target}}$, acting on $\mathbb{C}^{K}$.
For this purpose, we identify $K$ modes by choosing $\lfloor \frac{K}{2}\rfloor$ positions $\{0, 2, \cdots, 2\lfloor \frac{K}{2} \rfloor -2\}$ that are populated by both coin modes resulting in the mode set $Z = f(\{\ket{0}, \ket{2}, \cdots, \ket{2\lfloor \frac{K}{2}\rfloor -2}\} \otimes \{\ket{0}, \ket{1}\}) = \{\ket{0}, \ket{1}, \cdots, \ket{2\lfloor \frac{K}{2}\rfloor - 1}\}$ with $|Z| = 2\lfloor \frac{K}{2}\rfloor$.
This choice results in boundary cases, when one of the selected modes $z \in Z$ overlaps with a non-selected mode $m\in f(\mathcal{H}_e) \backslash Z$ in time ($\lfloor\frac{z}{2}\rfloor = \lfloor\frac{m}{2}\rfloor$).
In order to keep the selected subset of modes separated, we impose reflective boundary conditions by implementing the Pauli-X operation using the coin operation at these time bins.
The resulting beam splitter operation $\hat{T}_{z,m}$ implements the identity and will therefore be ignored in the following for the programming of the network.

\begin{figure}
    \includegraphics[width=0.45\textwidth]{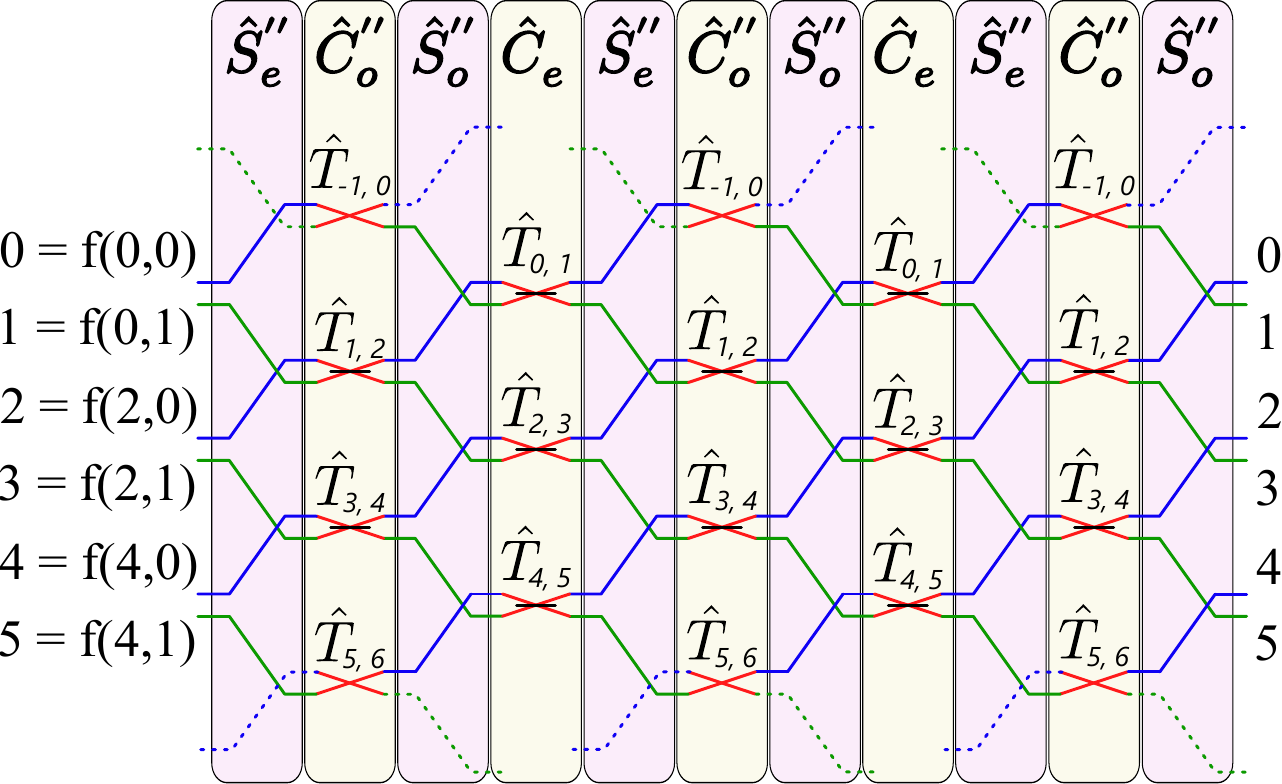}
    \caption{\label{fig:Interferometer} Schematics of an equivalent local network interferometer for a six-step quantum walk implementing an arbitrary $6 \times 6$ unitary. Note that the resulting beam-splitter operations $\hat{T}_{-1, 0}$ and $\hat{T}_{5, 6}$ implement a fixed Pauli-X gate realizing the required boundary conditions.
    }
\end{figure}

An example of the resulting network structure for a $6\times6$ target unitary is displayed in Fig. \ref{fig:Interferometer}.
The sequence of beam-splitter operations that defines our quantum walk evolution unitary is
    $
        \hat{U}_N =
    $ $
        \hat{T}_{1, 2}(0) \hat{T}_{3, 4}(0)
    $ $
        \hat{T}_{0, 1}(1) \hat{T}_{2, 3}(1) \hat{T}_{4, 5}(1)
    $ $
        \hat{T}_{1, 2}(2) \hat{T}_{3, 4}(2)
    $ $
        \hat{T}_{0, 1}(3) \hat{T}_{2, 3}(3) \hat{T}_{4, 5}(3)
    $ $
        \hat{T}_{1, 2}(4) \hat{T}_{3, 4}(4) =\hat{U}_{\mathrm{Target}}$,
where we already filtered out the coin operations like $\hat{T}_{-1, 0}(0)$ that implement Pauli-X in order to realize the boundary condition and therefore cannot be used to program the unitary.

\subsection{Programming beam splitters}

So far, we have decomposed a quantum walk into a specific sequence of local beam-splitter operations $\hat{T}_{a,b}$.
The next step is to determine the values of reflectivities and phases of each beam splitter for the specific target unitary.
As a quantum walk maps to a local network of beam splitters, we can follow the proposed protocol in Ref \cite{debrugière2025newdesignslinearoptical}.
They show that programming a local network to implement an arbitrary unitary is equivalent to sorting a list.
Specifically, they show that one can map any target unitary $\hat{U}_K \in \mathbb{C}^{K \otimes K}$ of arbitrary size $K$ to its list of mode indices $l=\sigma(1, 2, \dots, K)$ and a bijective permutation $\sigma$.
From there, programming the local network is equivalent to sorting this list, where each beam splitter allows one to swap two elements in the list.
Once the list is sorted from lowest to highest mode index $(1, 2, \dots, K)$, the unitary is programmed and the compiler is completed. 
For more details, see Sec. \ref{sec:LocalNetworkProgramming}.

As derived above, our local network is comprised of the beam splitter sequence $M_{2n+1} = (\hat{T}_{-1, 0}, \cdots, \hat{T}_{2\left\lfloor \frac{K}{2} \right\rfloor - 1, 2\left\lfloor \frac{K}{2} \right\rfloor})$ in odd steps and $M_{2n} = (\hat{T}_{0, 1}, \cdots, \hat{T}_{2\left\lfloor \frac{K}{2} \right\rfloor, 2\left\lfloor \frac{K}{2} \right\rfloor + 1})$ in even steps.
This effectively means that every even step we can swap the modes $(j - 1, j)$, while every odd step we can swap the modes $(j, j + 1)$ for $j \in \{0, 2, 4, \dots, 2\left\lfloor \frac{K}{2} \right\rfloor\}$.
This matches exactly the sequence of operations required for the sorting algorithm known as parallel bubble sort \cite{habermann1972parallel}.

Therefore, programming our system to implement an arbitrary unitary operation is equivalent to performing parallel bubble sort on a list.
As parallel bubble sort is capable of sorting any permuted list, we have thus proven that our system is capable of implementing any unitary.
Secondly, as an inverted list $l=(K, K-1, \dots, 1)$ is the worst case scenario for the algorithm in which it requires at most $N (N - 1) / 2$ operations and we can map any unitary to an unsorted list, we have shown that at maximum we require $N (N - 1) / 2$ coin operations or $N$ quantum walk steps.
Furthermore, this means that for any other permutation, only a subset of the full $N$-step quantum walk is required. 
Note that there exists a mathematical decomposition of the target matrix, called the Bruhat decomposition, which decomposes any given matrix into its list of mode indices $l=\sigma(1, 2, \dots, K)$ and all possible permutations $\sigma$.
Therefore, this decomposition can potentially find a shorter compile time and shallower quantum walk circuit for any target unitary, as demonstrated in \cite{debrugière2025newdesignslinearoptical}.

\section{Compiler Protocol \label{sec:CompilerProtocol}}

As we have seen above, a quantum walk with a two-dimensional coin can be represented by repeatedly applying a sequence of beam splitters and routing operations.
In the following, we present an algorithm to determine the values of the coin operations that need to be implemented so that the quantum walk compiles to the target $K \times K$ matrix, $\hat{U}_{\mathrm{Target}}$.\\

\textbf{Step 0: } We select $\left\lfloor \frac{K}{2} + 2 \right\rfloor$ even numbered positions $\{-2, 0, 2, 4, \cdots, 2\left\lfloor \frac{K}{2} \right\rfloor\}$.
As each position has two coin modes, this results in $2\left\lfloor \frac{K}{2} + 2 \right\rfloor$ modes involved in the quantum walk evolution.
We label each mode given by the position $x$ and the coin state $p$ as $z=x + p$.
This results in the set of modes $Z = \{-2, -1, 0, \cdots, 2\left\lfloor \frac{K}{2} \right\rfloor + 1\}$. Note that, in general, the compiler requires $K$ quantum walk steps in order to implement the desired target unitary.\\

\textbf{Step 1: } We collect the beam splitter sequence $M_n$ for each quantum walk step $n \in \{1, \cdots, K\}$.
\begin{itemize}
    \item If $n$ is odd, we retrieve the beam splitter sequence $M_n = (\hat{T}_{-1, 0}, \hat{T}_{1, 2}, \cdots, \hat{T}_{2\left\lfloor \frac{K}{2} \right\rfloor - 1, 2\left\lfloor \frac{K}{2} \right\rfloor})$.
    \item If $n$ is even, we retrieve the beam splitter sequence $M_n = (\hat{T}_{0, 1}, \hat{T}_{2, 3}, \cdots, \hat{T}_{2\left\lfloor \frac{K}{2} \right\rfloor, 2\left\lfloor \frac{K}{2} \right\rfloor + 1})$.
\end{itemize}

\textbf{Step 2: } We decompose our target unitary $\hat{U}_{\mathrm{Target}}$ into the two upper diagonal matrices $\hat{A}_1$ and $\hat{A}_2$ as well as the permutation operator $\hat{F}$ such that $\hat{U}_{\mathrm{Target}}^\dagger = \hat{A}_1\hat{F}\hat{A}_2$.
Next, we retrieve the unsorted sequence $l = (l_0, \cdots, l_{K - 1})$ from the permutation operator $\hat{F}$ such that each entry is $(F)_{j, l_j} = 1$.\\ 

\textbf{Step 3: } We perform the parallel bubble sort algorithm on the unsorted list $l$ using our quantum walk beam splitter sequence $M_n$. For each step $n \in \{1, \cdots, K\}$ and each beam splitter $\hat{T}_{m, m + 1} \in M_n$, we then proceed as follows:
\begin{itemize}
    \item If $m < 0$ or $m + 1 \geq K$, then we implement boundary conditions by setting the beam-splitter operation to identity;
    \item else if $l_{m} \leq l_{m + 1}$, then set the beam splitter-operation to identity;
    \item else if $l_{m} > l_{m + 1}$, then swap $l_{m}$ and $l_{m + 1}$. For this, we set the beam splitter parameters such that \begin{equation}
        \frac{(A_1)_{m, m + 1}}{(A_1)_{m + 1, m + 1}}
        =
        - \frac{T_{m,m + 1}^{(1, 1)}}{T_{m, m + 1}^{(1, 0)}}
        =
        (A_1)_{m, m + 1}.
    \end{equation}
    We further set $(\hat{A}_1^{'})_{m,m}, (\hat{A}_1^{'})_{m + 1, m + 1} = 1$ by propagating a diagonal matrix through to $\hat{A}_2$\cite{debrugière2025newdesignslinearoptical}.
\end{itemize}
With the sorted list, the permutation operator $\hat{F}$ implements the identity, and we can retrieve the diagonal operation $\hat{D} = \hat{A}_1^{'}\hat{A}_2^{'}$.
\\

\textbf{Step 4: } We reduce the footprint of the evolution by dropping all steps that implement only identity.\\

\textbf{Step 5: } We adjust all beam-splitter operations by applying a Pauli-X gate in the beginning, compensating for the permutation of the shift operator. Then, we read out each beam splitter $\hat{T}_{m, m + 1} \in M_n$ at each step $n$ as the coin operations that needs to programmed with $(\forall i, j \in \{0, 1\}) c_{i, j}^{(m)}(n) = T_{m, m + 1}^{(m + i, m + j)}(n)$.\\

\textbf{Step 6: } We perform the quantum walk evolution $\hat{U}_N$ as determined in steps zero to five, followed by phase shifts according to the diagonal operation $\hat{D}^\dagger$ on each of the resulting modes. The resulting unitary operation is $\hat{D}^\dagger \hat{U}_N = \hat{U}_{\mathrm{Target}}$.

\section{\label{sec:Setup} Time-Multiplexed Architecture}

\begin{figure*}
    \centering
    \includegraphics[width=0.9\textwidth]{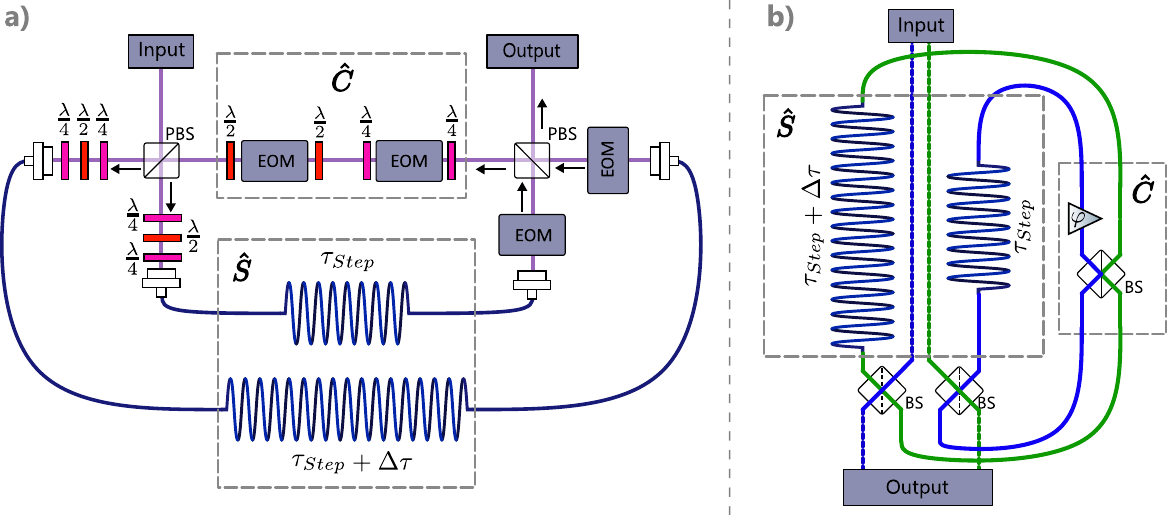}
    \caption{Experimental implementations of a time-multiplexed quantum walk, where panel a) shows a coin operation using polarization encoding, and panel b) shows a coin operation implemented using dual rail encoding. 
    Specifically, from a) to b), the electro-optic modulators (EOM) for the coin translate to an arbitrary beam splitter (BS) followed by an arbitrary phase shifter ($\varphi$), and the EOMs in the unbalanced Mach-Zehnder interferometer translate to two beam splitters that can be set to complete transmission or complete reflectance for the incoupling/outcoupling.
    \label{fig:Setup}}
\end{figure*}

In this section, we present an experimental scheme based on the time-multiplexed architecture \cite{PhysRevLett.104.050502} capable of implementing our proposed quantum-walk-based universal processor.
The complete setup is displayed in Fig. \ref{fig:Setup}.
Note that the structure of the experimental system has already been successfully used for many applications involving discrete-time quantum walks \cite{PhysRevLett.104.050502, Hamilton_2016, doi:10.1126/sciadv.aar6444, Pegoraro_2023}.
Because of the recursive structure of the quantum walk, it becomes intuitive to implement the processor in a loop architecture. 
We encode the two-dimensional coin degree of freedom in polarization, and the position degree of freedom is realized using time-bin encoding.
The coin operation must be programmed individually for each position $x$ at each step $n$, thus, a high-speed programmable polarization device is needed.
A suitable option would be an electro-optic modulator (EOM) operating in the megahertz regime using bulk crystals \cite{doi:10.1126/sciadv.adj0993} and gigahertz on integrated platforms \cite{Pan2022}.
Typically, EOMs induce a controllable phase shift in the circular polarization basis, resulting in the transformation matrix
\begin{equation}
    \hat{E}(\alpha) =
        \begin{pmatrix}
            \cos{\alpha} & i\sin{\alpha}\\
            i\sin{\alpha} & \cos{\alpha}
        \end{pmatrix},
\end{equation}
where the angle $\alpha \propto V$ has a linear dependence on the applied voltage $V$. 
Using quarter-wave plates $\hat{Q}(\beta)$, we can transform this operation such that it rotates the polarization along the equator of the Bloch sphere,
\begin{equation}
    \hat{Q}(0^\circ)\hat{E}\big(\alpha(x, n)\big)\hat{Q}(90^\circ)
            = \begin{pmatrix}
                \cos{\alpha(x, n)} & \sin{\alpha(x, n)}\\
                -\sin{\alpha(x, n)} & \cos{\alpha(x, n)}\\
            \end{pmatrix},
\end{equation}
while applying half-wave plates $\hat{H}(\beta)$ allows for an arbitrary phase shift,
\begin{equation}
    \hat{H}(22.5^\circ)\hat{E}(\phi(x, n))\hat{H}(22.5^\circ)
        = -e^{i\phi(x, n)} \begin{pmatrix}
            1 & 0 \\
            0 & e^{-2i\phi(x, n)}
        \end{pmatrix},
\end{equation}
at every temporally encoded step $n$ and position $x$.
Then, with two EOMs, one can construct an arbitrary beam splitter transformation of the form
\begin{equation}
    \begin{aligned}
        \hat{T}_{a,b}(x, n)
        = \begin{pmatrix}
            \ddots & \vdots & \vdots & \vdots & \vdots & \\
            \cdots & 1 & 0 & 0 & 0 & \cdots\\
            \cdots & 0 & \cos{\alpha}e^{-2i\phi} & \sin{\alpha} & 0 & \cdots\\
            \cdots & 0 & -\sin{\alpha}e^{-2i\phi} & \cos{\alpha} & 0 & \cdots\\
            \cdots & 0 & 0 & 0 & 1 & \cdots\\
            & \vdots & \vdots & \vdots & \vdots & \ddots\\
        \end{pmatrix},
    \end{aligned}
\end{equation}
where $\alpha=\alpha(x, n)$ and $\phi=\phi(x, n)$.
The voltage to apply to the EOMs at every step $n$ and every position $x$ is directly determined by the compiler algorithm presented in section \ref{sec:CompilerProtocol}.\\

The shift operator is performed by an unbalanced Mach–Zehnder interferometer, implemented by two polarizing beam splitters (PBS); see panel a) in Fig. \ref{fig:Setup}. 
The two polarization modes at each position enter one port of a PBS, which separates them into two optical paths with different lengths.
This effectively means that the time bin for one polarization is shifted by $\tau$, while the other is shifted by $\tau + \Delta \tau$.
Finally, a second PBS is oriented so that both optical paths are recombined and forwarded towards the coin operation if the polarization is unchanged in the arms of the unbalanced Mach-Zehnder interferometer.
Summing up, we can write the resulting operation explicitly as
\begin{equation}
    \hat{S}_{\tau} 
        = \sum_{t \in \Gamma} \ket{t + \tau + \Delta \tau}\bra{t} \otimes \ket{H}\bra{H} + \ket{t + \tau}\bra{t} \otimes \ket{V}\bra{V},
\end{equation}
where $\Gamma$ is the set of all time-bins.
If we now shift our reference frame after each shift operator by $\tau + \frac{\Delta \tau}{2}$, we retrieve the desired shift operator
\begin{equation}
    \hat{S}_{\mathrm{T}}
        = \sum_{t \in \Gamma} \ket{t + \frac{\Delta \tau}{2}}\bra{t} \otimes \ket{H}\bra{H} + \ket{t - \frac{\Delta \tau}{2}}\bra{t} \otimes \ket{V}\bra{V},
\end{equation}
up to the isomorphism $g: \Gamma \rightarrow \mathbb{Z}, (t) \stackrel{g}{\mapsto} t / \frac{\Delta \tau}{2}$ for $\Gamma = \{x \frac{\Delta \tau}{2} : x \in \mathbb{Z}\}$. 
If we now pump this system with a pulse train with a pulse separation of $\Delta\tau$, we can populate up to $\tau / \Delta\tau - 1$ position modes. Note that this architecture allows us to directly utilize squeezed light from a parametric down-conversion type-II process, where the signal and idler fields are encoded in polarization.

Lastly, to insert light in our quantum walk and measure its outcome, we perform the in-coupling into and out-coupling from the loop by switching the polarization inside the unbalanced Mach-Zehnder interferometer.
This allows us to choose the second port of the out-coupling PBS as the output, while the second port of the in-coupling PBS serves as the input.
Additional waveplates in each arm of the unbalanced Mach-Zehnder interferometer allow us to compensate for polarization rotations due to fiber-optical perturbations.

\section{Performance analysis}
\label{sec:Performance}
In this section, we investigate the effect of imperfections --- in particular, phase-noise and losses --- on the performance of the proposed system.
Then, we compare the resulting behavior with the performance of universal processors implemented using other prominent architectures based on spatial and temporal encoding.

\subsection{Phase and loss resilience}

We model experimental imperfections such as phase noise and photon loss using modified versions of the shift and coin operators
\begin{equation}
    \begin{aligned}
        \hat{\mathcal{S}}(n) &= \sum_{x \in \mathbb{Z}} \sum_{p = 0}^1 
            h(p, x, n) \ket{x - 1 + 2p}\bra{x} \otimes \ket{p}\bra{p},\\
        \hat{\mathcal{C}}(n) &= \sum_{x \in \mathbb{Z}} \sum_{p,q = 0}^1
            g(p, q, x, n) c_{p,q}^{(x)}(n) \ket{x}\bra{x} \otimes \ket{p}\bra{q},  
    \end{aligned}
    \label{eq:ImperfectionTerms}
\end{equation}
which are, in general, no longer unitary.
Here, $h(p, x,n) \in \mathbb{C}$ and $g(p, q, x, n) \in \mathbb{C}$ account for the experimental imperfections depending on the coin degree of freedom $p, q$, position $x$, and step $n$, and are referred to as imperfection terms henceforth.
Note that losses imply that these terms have an absolute value below one, and phase noise results in a non-zero complex phase.

To formally understand this model, we can consider the general Lindblad master equation governing the realistic process
\cite{L76,GK76}
\begin{equation}
    \partial_{t}\hat{\rho} = 
          \frac{1}{i\hbar}\left[\hat{H}, \hat{\rho}\right]
        - \sum_{x,p,n}
            \left(
                \frac{1}{2} \left\{ \hat{L}_{x,p,n}^\dagger \hat{L}_{x,p,n}, \hat{\rho} \right\} - \hat{L}_{x,p,n}^\dagger \hat{\rho} \hat{L}_{x,p,n}
            \right),
\end{equation}
where $[\,\cdot\,,\,\cdot\,]$ and $\{\,\cdot\,,\,\cdot\,\}$ represent the commutator and anti-commutator, respectively.
The unitary evolution of the system is governed by the Hamiltonian, $\hat{H}$, while dissipation at each step $n$, position $x$, and for each coin degree of freedom $p$ is captured through the Lindblad operators $\hat{L}_{x,p,n}$.
Effectively, our loss model considers the damping term $\frac{1}{2} \left\{ \hat{L}_{x,p,n}^\dagger \hat{L}_{x,p,n}, \hat{\rho} \right\}$, while neglecting the jumping $\hat{L}_{x,p,n}^\dagger \hat{\rho} \hat{L}_{x,p,n}$.
This is a reasonable approximation if the probability for a quantum jump is efficiently small \cite{Ju2019}.
Note that the solution where only the damping term is considered resembles a post-selection procedure, fixing the input and output photon numbers, and results in $|g(x,p,n)|, |h(x,p,n)| < 1$ in Eq. (\ref{eq:ImperfectionTerms}).
Consequently the following derivation cannot be directly applied to experiments such as Gaussian boson sampling \cite{PhysRevLett.119.170501} if losses are present, but unleashes its potential in quantum information processing protocols with single photons.

As we investigate a time-multiplexed system that completes $N$ steps in the realm of microseconds \cite{PhysRevLett.104.050502}, we can assume that temporal and acoustic fluctuations are sufficiently small to not influence our evolution.
This means that our imperfections are time-insensitive for each run, thus, independent of the position $x$ and constant for all steps $n$.
We can therefore reduce the dependence of our imperfection terms just to the coin degree of freedom, $h(p)$ and $g(p,q)$.

In addition, we assume a separable imperfection term $g(p,q) = g_o(p)g_i(q)$, where $g_{o/i}$ represents arbitrary phase shifts and losses before ($i$) and after ($o$) the ideal operation.
This allows us to construct the diagonal operators $\hat{G}_{i/o} = \hat{\mathbbm{1}}_{\mathrm{pos}} \otimes \sum_{p = 0}^1 g_{i/o}(p) \ket{p}\bra{p}$, only acting on the coin degree of freedom and thus commuting with the shift operation.
In this manner, the non-ideal coin operator $\hat{\mathcal{C}}(n)$ can be related to the ideal operator $\hat{C}(n)$ by means of diagonal operations,
\begin{equation}
    \hat{\mathcal{C}}(n) = \hat{G}_{i} \hat{C}(n) \hat{G}_{o}.
\end{equation}

As the next step, we look for a similar relation between the ideal and non-ideal shift operators in terms of diagonal matrices.
To this end, we reformulate the non-ideal shift operator as
$
    \hat{\mathcal{S}} = \sum_{x, y \in \mathbb{Z}} \sum_{p = 0}^1
        h(p) \delta_{y, x - 1 + 2p} \ket{y}\bra{x} \otimes \ket{p}\bra{p}
$
where we introduce the Kronecker delta specifying that we only evaluate our function where $y = x - 1 + 2p$.
Using this Kronecker delta, we can reformulate our imperfection term in terms of the positions $x$ and $y$ as
\begin{equation}
    h(p) \delta_{p, (y + 1 - x)/2} = \underbrace{\sqrt{h(0)h(1)}}_{g} {\underbrace{\left(\sqrt{\frac{h(0)}{h(1)}}\right)}_{k}}^{y - x},
\end{equation}
where we can identify a global term $g \in \mathbb{C}$ and a coefficient $k \in \mathbb{C}$ as our imperfection term.
Using this definition, we can write the non-ideal shift operator as
\begin{equation}
    \begin{aligned}
        \hat{\mathcal{S}}
            = g \hat{D} \hat{S} \hat{D}^{-1},
    \end{aligned}
\end{equation}
where $\hat{D} = \sum_{z \in \mathbb{Z}} k^z \ket{z}\bra{z} \otimes \hat{\mathbbm{1}}_{\mathrm{coin}}$ is a diagonal matrix.
This shows that the non-ideal shift operator can be written as the ideal shift operator plus diagonal operations and a global term.
Note that the diagonal operator acts only on the position subsystem and therefore commutes with the coin operation.

Putting both the non-ideal shift and coin operator together, we can construct the corresponding non-ideal, non-unitary evolution operator at every step $n$, $\hat{\mathcal{U}}(n) = \hat{\mathcal{C}}(n)\hat{\mathcal{S}}$.
After $N$ steps, the non-ideal evolution operation $\hat{\mathcal{U}}_N$ reads as
\begin{equation}
\label{eq:NoisyEvolution}
    \begin{aligned}
        \hat{\mathcal{U}}_N 
            &= \prod_{n=0}^N \big[
                \hat{\mathcal{U}}(n)
            \big]
            = \hat{G}_{i} \prod_{n=0}^N \big[
                \hat{C}(n) \underbrace{\hat{G}_{o} \hat{\mathcal{S}} \hat{G}_{i}}_{\hat{\mathcal{S}}^{'}}
            \big] \hat{G}_{i}^{-1}\\
            &= \hat{G}_{i} \prod_{n=0}^N \big[
                \hat{C}(n) g \hat{D} \hat{S} \hat{D}^{-1}
            \big] \hat{G}_{i}^{-1}\\
            &= g^N \underbrace{\hat{G}_{i} \hat{D}}_{\hat{A}} \prod_{n=0}^N \big[
                \hat{C}(n) \hat{S}
            \big] \underbrace{\hat{D}^{-1} \hat{G}_{i}^{-1}}_{\hat{A}^{-1}}
            = g^N \hat{A} \hat{U}_N \hat{A}^{-1},
    \end{aligned}
\end{equation}
consisting of the ideal operation $\hat{U}_N$ plus an imperfection dependent diagonal matrix $\hat{A} \in \mathbb{C}^{N \times N}$ and global term $g^N \in \mathbb{C}$.
From the explicit form of the imperfection matrix
\begin{equation}
    \hat{A} 
        = \sum_{x \in \mathbb{Z}} \left(\frac{h^{'}(0)}{h^{'}(1)}\right)^{z/2} \ket{z}\bra{z}
            \otimes \sum_{p = 0}^1 g_i(p) \ket{p}\bra{p}
\end{equation}
with $h^{'}(p) = h(p)g_i(p)g_o(p)$, we see that the experimental noise impacts the time bins $z$ independently from the coin modes $p$.
Under typical experimental conditions, we expect that the coin-noise process effects both polarizations nearly identically ($g_i(0) \approx g_i(1)$) as both polarizations share a common spatial path.
With these considerations, the strength of our hybrid encoding consisting of multiple degrees of freedom becomes apparent as the expected impact of experimental noise is near zero on the coin subsystem.
This allows us, for example, to easily perform partial state tomography independently of the experimental noise, measuring the coin state at each position.\\
We further analyze the implications of the diagonal noise matrices $\hat{A}$ on the implemented unitary evolution depending on the individual contributions.

If the considered imperfections only amount to a phase difference ($h^{'}(0)=e^{i\alpha}, h^{'}(1)=e^{i\beta}$), this results in the coefficient term $e^{i\frac{\alpha-\beta}{2z}}$.
The resulting diagonal matrices apply a phase shift before and after the desired interferometer.
In the framework of boson sampling, this means that for scattershot boson sampling \cite{PhysRevLett.113.100502}, these diagonal matrices are of no consequence, while they change the considered unitary transformation for Gaussian boson sampling \cite{PhysRevLett.119.170501}.

If the considered imperfections only amount to a loss term ($h^{'}(0)=a, h^{'}(1)=b$), this results in the coefficient term $k=\sqrt{a/b}$.
Note that, for balanced losses $a=b$, $k=1$ and the diagonal matrix $\hat{D}$ becomes the identity.
Unbalanced losses, however, manifest themselves as a shift of the probability at the input and output of the unitary, impacting both scattershot and Gaussian boson sampling.

\subsection{Comparison between architectures}

To conclude, we compare the performance of our proposed scheme with other known architectures in terms of fault tolerance. 
We chose the architectures proposed by Reck et. al. \cite{PhysRevLett.73.58} and by Clements et. al. \cite{Clements:16} as the main candidates for spatial implementations, and the MGDR architecture \cite{ PhysRevLett.113.120501} as an example of a different time-multiplexed setup.

All of the above schemes decompose the target unitary into a sequence of individual beam-splitter operations between neighboring optical modes. 
Therefore, imperfections mainly arise in terms of losses and phase shifts from the connection between individual beam splitters. 
Accordingly, we use the imperfection model where the individual beam splitters are considered ideal, while the connections introduce the experimental imperfection. 
This results in the beam-splitter transformation
\begin{equation}
    \hat{T}_{a,b} \mapsto
        \begin{pmatrix} \alpha^{(a,b)}_o & 0 \\ 0 & \beta^{(a,b)}_o \end{pmatrix}
        \hat{T}_{a,b}
        \begin{pmatrix} \alpha^{(a,b)}_i & 0 \\ 0 & \beta^{(a,b)}_i \end{pmatrix},
\end{equation}
where $\alpha^{(a,b)}_{o/i}, \beta^{(a,b)}_{o/i} \in \mathbb{C}_{\leq 1}$ are arbitrary parameters representing loss and phase shifts at the input ($i$) and output ($o$) of a beam splitter for each mode $a,b$.
Note that, for time-multiplexed systems, these imperfection parameters are assumed to be constant over all modes as the same physical components are reused in a loop during the evolution for all modes.

We characterize the effect of these imperfections by averring the fidelity $\mathcal{F}$ and similarity $\mathcal{O}$ over five hundred $20\times20$ unitaries, each with five hundred random patterns of imperfections chosen within a normal distribution with a standard deviation of $\sigma$. 
We define the fidelity as
\begin{eqnarray}
    \mathcal{F} = \left|\frac{
        \mathrm{tr}\left(
            \hat{\mathcal{U}}_N \hat{U}_{\mathrm{Target}}^\dagger
        \right)}{\sqrt{N} \sqrt{\mathrm{tr}\left(
            \hat{\mathcal{U}}_N\hat{\mathcal{U}}_N^\dagger
        \right)}}\right|^2,
\end{eqnarray}
where $\hat{\mathcal{U}}_N$ is the resulting transformation matrix including experimental imperfections. 
This standard measure of fidelity for unitaries disregards global terms, while still being sensitive to the relative phase of the individual transition amplitudes. 
On the other hand, the similarity defined as
\begin{equation}
    \mathcal{O} = \frac{
        \sum_{x,y=0}^N \left| 
            \big(\mathcal{U}_N\big)_{x,y} \big(U_{\mathrm{Target}}\big)_{x,y} 
        \right|}{
            \sqrt{\sum_{x,y=0}^N \left|\big(\mathcal{U}_N\big)_{x,y}\right|^2}
            \sqrt{\sum_{x,y=0}^N \left|\big(U_{\mathrm{Target}}\big)_{x,y}\right|^2}
        },
\end{equation}
where we treat the individual operators as high-dimensional vectors and then calculate their overlap, is insensitive to all phase terms.

\begin{figure}[h]
    \centering
    \includegraphics[clip,width=0.46\textwidth]{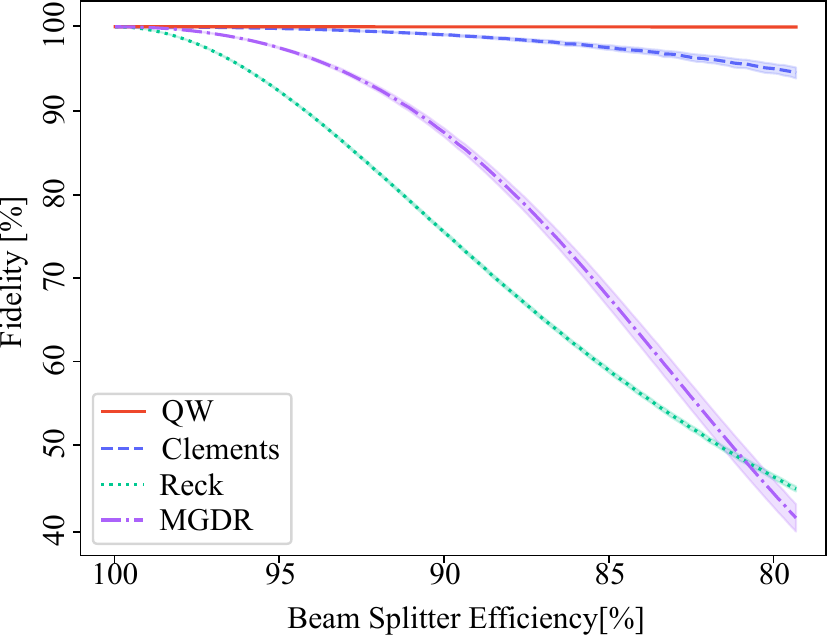}  
    \caption{Effect of randomly varying beam-splitter efficiencies with a standard deviation of $\sigma=5\%$ on the fidelity with the target matrix. The shaded area represents one standard deviation over all averaged unitaries and loss patterns.}
    \label{fig:LossFidelity}
\end{figure}

We first study the effect of losses per beam splitter on the resulting fidelity for all four architectures by sampling the imperfection terms from a Gaussian distribution. 
The results are displayed in Fig. \ref{fig:LossFidelity} for varying losses with $\sigma = 5\%$ per beam splitter.

The first observation we can make is that the fidelity of the implemented unitary in the quantum walk architecture is completely independent of the average loss.
In contrast, the Clements architecture becomes marginally affected, and, for the Reck and the MGDR architectures, the fidelity drops significantly.
This difference in the effect of losses on the resulting fidelity comes from the symmetrization of the arrangement of the constituting operations.
As discussed in Ref. \cite{Clements:16}, unbalanced loss resulting from unsymmetrical arrangements has a non-recoverable impact on the resulting interference.
Therefore, architectures such as the Reck or MGDR are strongly affected, while more symmetrical arrangements such as the Clements architecture are only marginally affected.
Note that the Clements architecture is affected due to its border conditions, while our system implements border conditions indistinguishably from the coin operations, which result in complete symmetry of the underlying local network.

\begin{figure}
    \centering
        \includegraphics[width=0.46\textwidth]{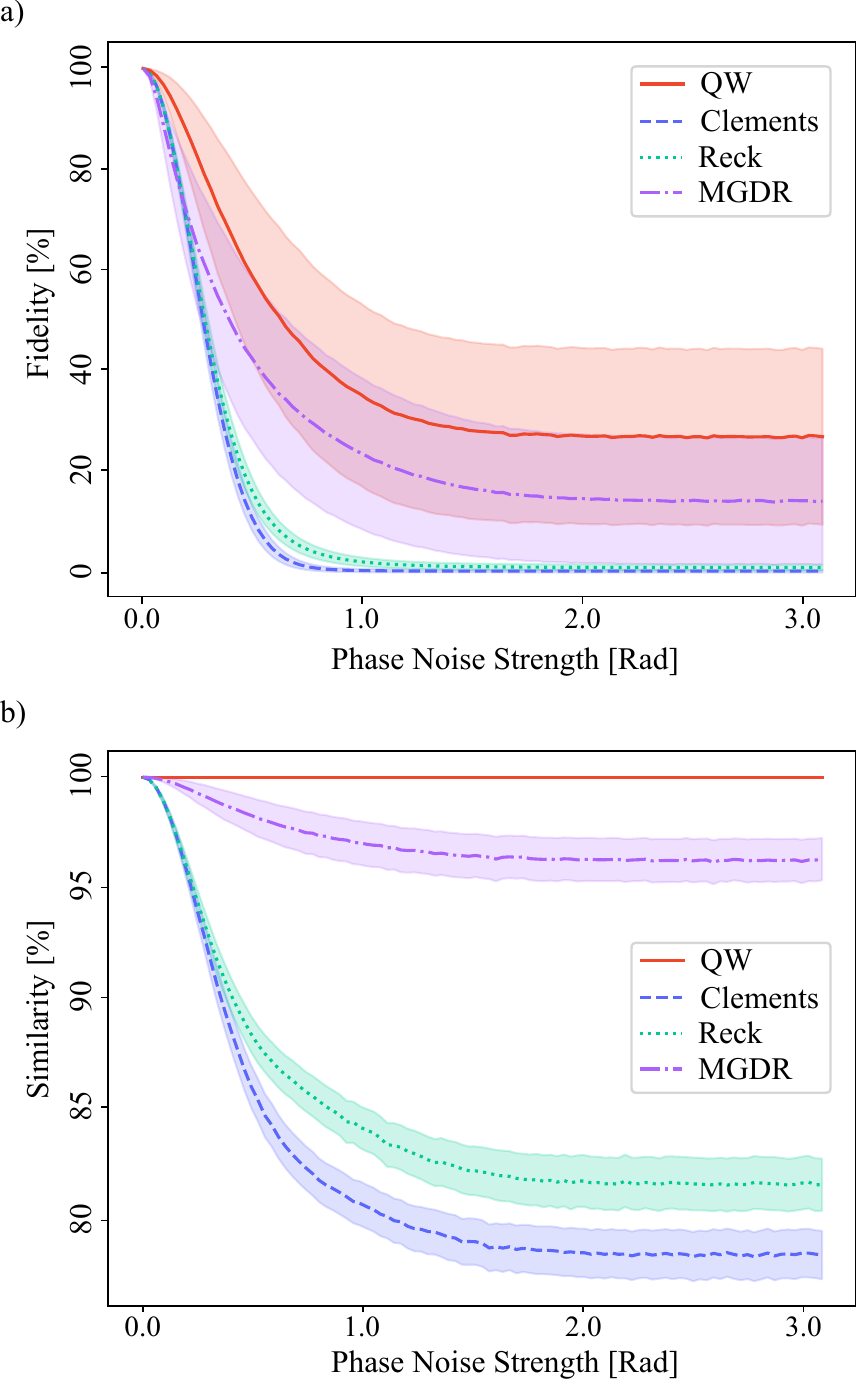}
        \label{fig:PhasenoiseFidelity}
        

    \caption{Effect of random phase noise on the resulting fidelity, panel a),  and similarity, panel b), with the target matrix. The shaded area represents one standard deviation over all averaged unitaries and phase noise patterns. }
    \label{fig:Phasenoise}
\end{figure}

Next, we investigate the effect of random phase shifts between individual beam splitters, with imperfection terms $|\alpha_{o/i}^{(a,b)}|^2 = |\beta_{o/i}^{(a,b)}|^2 = 1$.
For this purpose, we sample each imperfection term individually by randomly selecting a phase $e^{i\varphi}$ from a Gaussian distribution with varying standard deviations.
The results are displayed in Fig. \ref{fig:Phasenoise} for all four considered architectures.

Here, we can see that the similarity remains at one hundred percent, meaning that the underlying transition amplitudes of the desired unitary can be implemented in our quantum walk architecture completely independent of phase noise.
We further see that this not true for all other considered architectures, where such phase noise would need to be minimized.
Finally, we see that for the considered phase noise the resulting fidelity of the implemented unitary compares favorably with all other architectures.

\section{Conclusions}
\label{sec:Conclusion}

In this paper, we systematically investigated the performance of a coined discrete-time quantum walk as a photonic quantum processor.
Specifically, we show that for a two-dimensional coin space and reprogrammable coin operation, any $N \times N$ target unitary evolution can be realized in an $N$-step quantum walk.
This closes the gap between theory and experiments by providing a recipe to translate arbitrary unitaries into the experimental parameters required to be implemented in a quantum walk.
We proposed a time-multiplexed hybrid encoding where both the coin and position degree of freedom are exploited, requiring $\lfloor \frac{N(N-1)}{2} \rfloor$ time-bins and operations, as well as a fixed number of optical components for reproducing any $N \times N$ unitary evolution.
This makes the proposed system strongly scalable and more resource efficient in comparison to existing spatial \cite{PhysRevLett.73.58, Clements:16} and time-multiplexed \cite{PhysRevLett.113.120501, Yu2023} architectures.

Furthermore, we introduced an experimental setup capable of implementing our findings and provided the compiler algorithm that translates the target evolution into the individual coin operations to be programmed.
Our proposed architecture is capable of implementing two independent unitary evolutions in parallel, which could be beneficial for applications such as indefinite causal order experiments \cite{PhysRevLett.113.250402} or error correction protocols such as unitary averaging \cite{PhysRevA.109.062436}.
We further analyzed the impact of the most common experimental imperfections on the performance of the photonic quantum processor, proving that the resulting evolution of our quantum walk is completely insensitive to photon losses and highly resilient against phase noise.
Moreover, we showed that it compares favorably across multiple configurations of loss and phase noise with some of the most prominent existing architectures.

\section{Acknowledgments}

We acknowledge financial support by the European Commission through the Horizon Europe project EPIQUE (Grant No. 101135288).
We acknowledge the financial support of the
Deutsche Forschungsgemeinschaft (DFG) via the TRR 142/3 (Project No. 231447078, Subproject No. C10).

\section{Author Contributions}

J.L., P.H., and F.P. developed the experimental platform. J.L. with assistance of L.A. derived the protocol and theoretical proofs. J.L. and L.A. wrote the manuscript. P.H. contributed with ideas about the systems stability. C.S., J.S., and B.B. supervised the work. C.S. initiated the work. All authors discussed the results and commented on the manuscript.

\appendix

\section{\label{sec:LocalNetworkProgramming} Local network programming}

For completeness, we provide a short summary of the result of T. G. de Brugière, et. al. \cite{debrugière2025newdesignslinearoptical} on the programming of a local network to compile as an arbitrary unitary. Further, we provide a method to represent any target unitary as an inverted list in their formalism as a worst-case compiler for our photonic processor.

First, we decompose the hermitian conjugate of our target matrix into two upper triangular matrices $\hat{A}_1, \hat{A}_2$ and two permutation matrices $\hat{P}, \hat{F}$, such that
\begin{equation}
    \hat{U}_{\mathrm{Target}}^\dagger = \hat{P} \hat{A}_1 \hat{F} \hat{A}_2.
    \label{eq:DecompositionPUFU}
\end{equation}
This decomposition can be achieved by, for example, using the LU-Decomposition decomposing $\hat{U}_{\mathrm{Target}}^\dagger$ into a lower triangular $\hat{L}_1$, upper triangular $\hat{A}_2$ and permutation $\hat{P}$ matrix, such that $\hat{U}_{\mathrm{Target}}^\dagger = \hat{P} \hat{L}_1 \hat{A}_2$.
From there we can transform the lower triangular matrix $\hat{L}_1 = \hat{F}^{-1}\hat{F}\hat{L_1}\hat{F}^{-1}\hat{F}$ using the anti diagonal operator $\hat{F}$, such that $\hat{U}_{\mathrm{Target}}^\dagger = \hat{P}^{'} \hat{A}_1 \hat{F} \hat{A}_2$, where $\hat{A}_1 = \hat{F}\hat{L_1}\hat{F}^{-1}$ and $\hat{P}^{'} = \hat{P} \hat{F}^{-1}$.

Given this representation, the next step is to investigate the impact of a local beam-splitter operation $\hat{T}_{a, a + 1}$ on our target matrix. Specifically, we consider
\begin{equation}
    \hat{T}_{a, a + 1} \hat{P}^\dagger \hat{U}_{\mathrm{Target}}^\dagger 
        = \underbrace{\hat{T}_{a, a + 1} \hat{A}_1 \hat{B}}_{\hat{A}_1^{'}} \underbrace{\hat{B}^{-1} \hat{F}}_{\hat{F}^{'}} \hat{A}_2,
    \label{eq:BeamsplitterOperationTUBBPUU}
\end{equation}
where $\hat{B}$ is an invertible matrix chosen such that $\hat{A}_1^{'}$ is again an upper triangular matrix and $\hat{F}^{'}$ a permutation matrix. 
First of all, if the beam splitters $\hat{T}_{a, a + 1}$ implements identity, nothing changes, and the resulting permutation operator $\hat{F}^{'} = \hat{F}$ remains unchanged. 
It has been proven \cite{debrugière2025newdesignslinearoptical} that the reflectivity and phase of any beam splitter $\hat{T}_{a, a + 1}$ can be chosen such that $\hat{B}$ implements a flip between mode $a$ and $a + 1$, and identity everywhere else.
This causes a flip of the row $a$ and $a + 1$ in the resulting permutation operator $\hat{F}^{'} = \hat{B}_1^{-1} \hat{F}$.
For any sequence of beam splitters $(T_{m,m+1}, m \in M)$ representing a local network $M$, which reduces the permutation operator $\hat{F}^{'} \mapsto \identity$ to identity, we see that
\begin{equation}
    \left(\prod_{m \in M} \hat{T}_{m,m + 1}\right) \hat{P}^\dagger \hat{U}_{\mathrm{Target}}^\dagger 
        = \hat{A}_1^{'} \hat{\mathbbm{1}} \hat{A}_2^{'} = \hat{D},
\end{equation}
where $\hat{D} = \hat{A}_1^{'} \hat{A}_2^{'}$ is a unitary upper triangular matrix and therefore a diagonal matrix.
From there, we can directly write our target matrix
\begin{equation}
    \hat{D}^\dagger \left(\prod_{m \in M} \hat{T}_{m, m + 1}\right) \hat{P}^\dagger = \hat{U}_{\mathrm{Target}},
\end{equation}
as this sequence of beam splitters up to a diagonal matrix $\hat{D}$ and permutation matrix $\hat{P}^\dagger$.

\begin{figure}[h]
    \centering
    \includegraphics[width=0.48\textwidth]{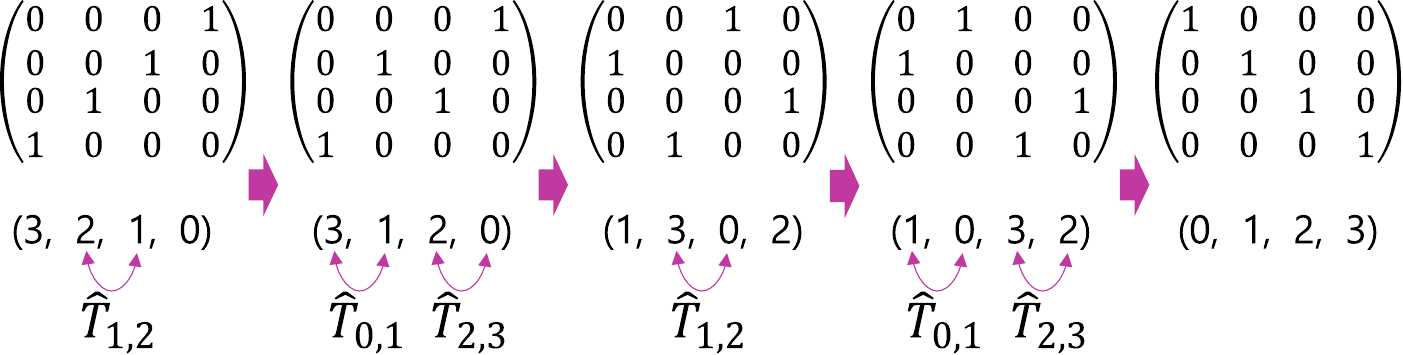}
    \caption{%
    Example process of programming an arbitrary $4\times4$ unitary using the resulting quantum walk beam-splitter sequence. \label{fig:Sorting}} 
\end{figure}

The next step is to consider the local network required in order fully transition the permutation operator $\hat{F}$ to the identity operation via local swaps.
Here, we map the permutation operator $\hat{F}$ resulting from the decomposition of an arbitrary target matrix $\hat{U}_{\mathrm{Target}}$ into the list of modes $l = \sigma(1, 2, \cdots, N)$ permuted by the bijective permutation $\sigma$.
Specifically, the permutation $\sigma$ maps each mode index $j$ to the index $\sigma(j) = l_j$ where the permutation operator $\hat{F}$ is nonzero $\big((F)_{j, l_j} = 1\big)$.
For example, the anti-diagonal permutation operator results in the sequence $l = (N - 1, N - 2, \cdots, 1, 0)$, while the identity operator results in $l = (0, 1, \cdots, N - 2, N - 1)$.
Now, we take the beam-splitter sequence $M$ and, for each beam splitter $\hat{T}_{m, m + 1}$ with $m \in M$, either flip the entry $l_m$ and $l_{m+1}$ of the sequence $l$ or keep it as it is.
See Fig. \ref{fig:Sorting} for a scheme of this process.
Once the resulting sequence $l$ is equal to the sequence for identity, the permutation operator $\hat{F}$ is equal to identity, which means that our chosen sequence of beam splitters $M$ can implement the target unitary $\hat{U}_{\mathrm{Target}}$.
Therefore, verifying that a sequence of beam splitters is capable of implementing the target matrix is equivalent to proving that it is capable of sorting the sequence $l$.

\bigskip

\bibliography{main}

\end{document}